\documentclass[preprint2]{aastex}
\usepackage{psfig}

\singlespace

\newcommand{\beq}{\begin{equation}}
\newcommand{\eeq}{\end{equation}}
\newcommand{\eeqn}[1]{\label{#1}\end{equation}}
\newcommand{\dt}[1]{\frac{\partial  #1}{\partial t}}
\newcommand{\eq}[1]{(\ref{#1})}
\newcommand{\vO}{{\bf \Omega}}
\newcommand{\na}{ {\bf \nabla} }
\newcommand{\vv}{{\bf v}}
\newcommand{\vu}{{\bf u}}
\newcommand{\vn}{{\bf n}}
\newcommand{\ez}{{\bf e}_z}
\newcommand{\ddz}[1]{\frac{\partial^2  #1}{\partial z^2}}
\renewcommand{\char}{characteristic}
\newcommand{\vzero}{{\bf 0}}
\newcommand{\bu}{\overline{u}}
\newcommand{\lp}{ \left(}
\newcommand{\rp}{ \right)}
\newcommand{\tu}{\tilde{u}}
\newcommand{\cth}{ \cos\theta }
\newcommand{\beqa}{\begin{eqnarray*}}
\newcommand{\beqan}{\begin{eqnarray}}
\newcommand{\eeqa}{\end{eqnarray*}}
\newcommand{\eeqan}[1]{\label{#1}\end{eqnarray}}
\newcommand{\sth}{ \sin\theta }
\newcommand{\lac}{ \left\{}
\newcommand{\rac}{ \right\}}
\newcommand{\lc}{ \left[}
\newcommand{\rc}{ \right]}
\newcommand{\dzeta}[1]{\frac{\partial  #1}{\partial \zeta}}
\newcommand{\intvol}{ \int_{(V)}\! }

\begin{document}
\title{Ekman layers and the damping of inertial r-modes in a spherical
shell: application to neutron stars}

\author{Michel Rieutord}
\affil{Observatoire Midi-Pyr\'en\'ees and Institut Universitaire de France}
\affil{14 avenue Edouard Belin, 31400 Toulouse}
\email{rieutord@obs-mip.fr}
\date{\today}

%\maketitle

\begin{abstract}
Recently, eigenmodes of rotating fluids, namely inertial modes, have
received much attention in relation to their destabilization when
coupled to gravitational radiation within neutron stars. However, these
modes have been known for a long time by fluid dynamicists. We give
a short account of their history and review our present understanding
of their properties. 
Considering the case of a spherical container, we then give the exact
solution of the boundary (Ekman) layer flow associated with inertial
r-modes and show that previous estimations all underestimated the
dissipation by these layers. We also show that the presence of an inner
core has little influence on this dissipation. As a conclusion, we
compute the window of instability in the temperature/rotation plane for
a crusted neutron star when it is modeled by an incompressible fluid.
\end{abstract}


\section{Introduction}

Recently much work has been devoted to the study  of the rotational
instability of neutron stars resulting from a coupling between
gravitational radiation and the so-called ``r-modes" of a rotating star
\cite{anderss98,FM98,LOM98,KS99}.
Such an instability may indeed play a key role in the distribution of
rotation periods of neutron stars as well as it may be an important source
of gravitational radiation.

In this paper, we shall first clarify a point of history concerning
``r-modes" which are in fact a special class of inertial modes; we shall
then review their singular properties which have been clarified only
very recently in \cite{RV97}, \cite{RGV99} and \cite{RGV00b}. The last
section will present the analytical derivation of the damping rate of
inertial r-modes in a neutron star with a crust and/or a core through
the boundary layer analysis within the framework of newtonian theory.
We conclude on the stability of crusted neutron stars when modeled by
an incompressible viscous fluid in a rotating sphere.

\section{A short point of history}

The very first work on rotating fluids oscillations which are presently
known as {\em inertial modes} dates back to Thomson\footnote{later Lord
Kelvin} (1880)\nocite{T1880} who analysed the case of a fluid contained in
a cylinder. However, another impetus to the study of these oscillations
was given soon after by Poincar\'e's (1885) \nocite{Poinc1885} work on
the stability of rotating self-gravitating masses, a work applied to
MacLaurin spheroids by \cite{Bryan1888}\footnote{ but
see the recent rederivation by \cite{LI99}.} and later continued
by \cite{Cartan22} who christened the equation of inertial modes
as ``Poincar\'e equation". In these studies, however, the effect of
rotation is combined to the one of gravity through (for an incompressible
fluid) surface gravity waves. In fact, except for the work of Thomson,
investigations on the oscillations specific of rotating fluids seem
to have started with the work of \cite{Bj33} where they are
called ``elastoid-inertial oscillations" since conservation of angular
momentum makes axis-centered rings of fluid behave elastically; but see
\cite{Fultz59} or \cite{Ald67} for an account on this part of history. In
the sixties, much work has been devoted to these oscillations, mainly by
Greenspan who introduced the terminology of ``inertial oscillations". The
presently used denomination ``inertial modes" has been ``officially"
given by Greenspan's book \citep{Green69}.

However, inertial modes are somewhat too general for applications in some
specific domains like atmospheric sciences. In this field indeed motions
are essentially two dimensional and inertial modes may be simplified
into the well-known Rossby (or planetary) waves.

The introduction of r-modes by \cite{PaPr78} was quite unfortunate
since they associated eigenmodes of rotating fluids with a very special
class of inertial modes, namely purely toroidal inertial modes.  This lead
following authors to introduce weird names such as ``hybrid modes" or
``generalized r-modes" \citep{LF99} for describing the general class of
inertial modes. We therefore encourage authors to use,
as fluid dynamicists, inertial modes unless they discuss the very specific
r-modes.

\section{The present theory of inertial modes}

Inertial modes are a class of modes of oscillation of rotating fluids
which owe their existence to the Coriolis force. This force of inertia
has indeed a restoring action on perturbations of rotating fluids since
it insures the global  conservation of angular momentum. These modes
have many properties similar to those of gravity modes of stably
stratified fluids \citep{RN99}.

The dynamics of inertial modes may be appreciated when all other effects
are suppressed: no compressibility, no magnetic fields, no gravity,
etc... only an incompressible inviscid rotating (like a solid body)
fluid. In this case, perturbations of velocity $\delta \vv$ and
pressure $\delta P$ obey

\beq \dt{\delta\vv} + 2\vO\times\delta\vv = -\na\delta P,
\qquad \na\cdot\delta\vv =0 \eeqn{dimeq}
where $\vO$ is the angular velocity of the fluid. Concentrating on
time-periodic oscillations and choosing $(2\Omega)^{-1}$ as the time
scale, \eq{dimeq} can be written

\beq i\omega\vu+\ez\times\vu = -\na p, \qquad \na\cdot\vu=0\eeqn{eqmo}
with non-dimensional variables; $\omega$ is the non-dimensional (real)
frequency. When the velocity $\vu$ is eliminated in favor of the pressure
perturbation $p$, one is left with

\beq \Delta p -\frac{1}{\omega^2}\ddz{p} = 0 \eeqn{poincare}
which is known as Poincar\'e equation since \cite{Cartan22}. This
equation is remarkable in the fact that it is {\em hyperbolic spatially} since
$|\omega| \leq 1$ \citep{Green69}. As the solution of \eq{poincare} must
meet boundary conditions, namely $\vu\cdot\vn=0$, we see that inertial
modes are solutions of an ill-posed boundary value problem\footnote{Such
boundary conditions eliminate any distorsion of the surface due to the
fluid motion; for a free surface, these distorsion are surface gravity
waves (see Rieutord 1997\nocite{rieutord97}) but their inclusion (in
order to be more realistic) would not modify the ill-posed nature of
the problem.}. This property means that, in general, inertial modes are
singular; in other words they cannot exist physically if the fluid is
strictly inviscid.  These properties are detailed in \cite{RV97} and
\cite{RGV99}; to make a long story short, one may summarize the
situation as follows.

\begin{figure}[t]
\centerline{\psfig{figure=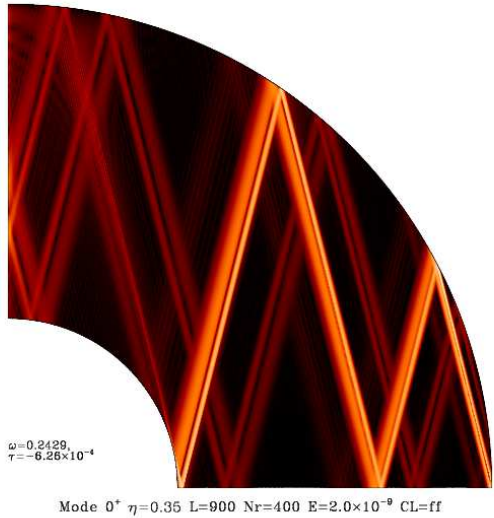,width=8cm}}
\caption[]{The kinetic energy distribution in a meridional plane of
an inertial mode in a spherical shell associated with an
equatorial attractor. A co-existing polar attractor is also slightly
excited. The mode is axisymmetric with equatorial symmetry. Stress-free
boundary conditions have been used on both shells; this solution
was computed with an Ekman number of 2 10$^{-9}$ and required 1300
spherical harmonics and 450 radial grid points (Gauss-Lobatto). The
ratio of the inner radius to the outer radius is $\eta=0.35$.
$\omega=0.2429$ and the damping rate $\tau=-6.26\times10^{-4}$ are given
in dimensionless units as equation~\eq{eqmo}.}

\label{in_attr} \end{figure}

Let us first recall that in hyperbolic systems, energy propagates along
the \char s of the equation. For the Poincar\'e problem, these are
straight lines in a meridional plane. A way to approach the solutions of
this difficult problem is to examine the propagation of \char s as they
reflect on the boundaries. They define trajectories which depend
strongly on the container. Let us therefore concentrate on the case of
a spherical shell as a container; this configuration is relevant for
neutron stars with a central core due to some phase transition of the
nuclear matter \cite[see][]{Haen96}. In this case, it may be shown that \char\
trajectories generically converge towards attractors which are periodic
orbits. It may be shown \citep{RGV99} that in this case, the associated
solutions are singular, namely the velocity field is not
square-integrable. However, inertial r-modes are still solutions of
the problem since they meet the boundary conditions ($u_r=0$); in fact
they are the only regular (square-integrable) solutions of the
Poincar\'e problem in a spherical shell. In a more mathematical way, we
may say that the spectrum of eigenvalues of the Poincar\'e problem in a
spherical shell is empty except for the inertial r-modes. In this
sense, these modes are quite exceptional. This situation occurs because
there exists no system of coordinate in which the dependent variables of
the Poincar\'e equation can be separated. This is a consequence of the
conflict between the symmetry of the Coriolis force (cylindrical) and
the geometry of the boundaries. Thus, when this constraint is relaxed,
like in the case of a cylindrical container, regular solutions exist
and a dense spectrum of eigenvalues appears in the allowed frequency
range, namely $[0,2\Omega]$. In the case the container is a full sphere,
attractors also disappear and eigenmodes exist; they are also related to a
dense spectrum of eigenfrequencies. In this case, Poincar\'e equation
is exactly solvable \citep{Green69}.

However, real fluids have viscosity ($\nu$) and equation \eq{eqmo}
should be transformed into

\beq \lambda\vu +\ez\times\vu = -\na p+E\Delta\vu, \qquad
\na\cdot\vu=0\eeqn{eqvis}
where $\lambda$ is the complex eigenvalue and $E=\nu/2\Omega R^2$ is the
Ekman number ($R$ is the outer radius of the shell).

Using no-slip ($\vu=\vzero$) or stress-free boundary conditions,
\eq{eqvis} yields a well-posed problem. Yet, the singularities of the
associated inviscid solutions show up through the existence of shear
layers. As shown by fig.~\ref{in_attr}, the shape of inertial modes is deeply
influenced by the underlying singularity of the inviscid solution. We
have shown \citep{RGV99} that these shear layers are in fact nested
layers with different scales for their inner part scales as
$E^{1/3}$ and their outer part seems to scale with $E^{1/4}$.
Because of these internal shear layers, these modes are strongly damped.

We therefore see that according to whether a neutron star has a central
core or not, the damping of inertial modes will be extremely different.
If there is a central core,  the only regular modes are the inertial
r-modes which will be by far the least damped; if there isn't any
core then a dense spectrum exists \citep{Green69,LF99} but inertial
r-modes  remain the most unstable because of their simple structure.

\section{Damping of toroidal inertial modes in a sphere or a shell}

We shall now give the expression of the viscous damping of
inertial r-modes when one of the boundary is solid therefore when the
dissipation is due to Ekman boundary layers; we shall thus complete the
works of \cite{BU00} and \cite{AJKS00} by giving the rigorous estimate
of the damping rate; the method which we use here has been outlined in
\cite{Green69}.

The damping rate is given by:

\beq \gamma = \Re e(\lambda)= -E\frac{\int (\na : \vu)^2 dV}{\int \vu^2 dV} \eeq
where $(\na : \vu)^2$ stands for the squared rate-of-strain tensor (see
below). The velocity field of r-modes is

\[ \bu_\theta = A r^m (\sin\theta)^{m-1}\sin(m\phi+\omega_m t) \]
\[ \bu_\varphi = A r^m (\sin\theta)^{m-1}\cos\theta\cos(m\phi+\omega_m t) \]

The kinetic energy integral may be evaluated explicitly

\beq \int \vu^2 dV = \pi A^2 \lp
1-\eta^{2m+3}\rp\frac{2^{m+1}(m+1)!}{m(2m+3)!!} \eeqn{ke}
where $\eta$ is the ratio of the radius of the inner boundary to the one
of the outer boundary.

The dissipation integral need more work if one of the boundary is
no-slip. In this case, dissipation is essentially coming from the Ekman layers
and thus we need to derive the flow in these layers. The method has been
given by Greenspan from whom we know that the boundary layer correction
$\tu$, is related to the interior solution $\bu$ by

\beq \tu_\theta+i\tu_\varphi = -(\bu_\theta+i\bu_\varphi)_{r=r_b}
e^{-\zeta\sqrt{i\cth\pm i\omega}} \eeq
where $\zeta$ is the radial scaled variable $(r-r_b)/\sqrt{E}$ with
$r_b$ as the radius of the boundary (1 or $\eta$). The complete solution is then $\vu=\tu+\bu$; setting $\beta=\omega t + m\varphi$  we have

\beqa \bu_\theta+i\bu_\varphi = \frac{A r^m (\sth)^{m-1}}{2i}\hspace*{2cm} \\
\qquad\times\lac (1-\cth)e^{i\beta} - (1+\cth)e^{-i\beta}\rac \eeqa
from which it follows that

\beqa \tu_\theta+i\tu_\varphi = \frac{A r_b^m (\sth)^{m-1}}{2i}
\lac (1+\cth)e^{-i\beta-\zeta\sqrt{i\cth-i\omega}}\right.\\ -
\left. (1-\cth)e^{i\beta-\zeta\sqrt{i\cth+i\omega}}\rac \eeqa

Now we need the expression of the square of the rate-of-strain tensor
$s_{ij}=\partial_iv_j+\partial_jv_i$ in spherical coordinates, viz

\[ (\na : \vu)^2= s_{rr}^2 +
s_{\theta \theta }^2 +s_{\phi\phi}^2 + 2(s_{r\theta }^2 +
s_{r\phi}^2 +s_{\theta \phi}^2) \]

Since the radial derivatives dominate, this expression reduces to the
contribution of the tangential stresses. Using the scaled coordinate,
$\zeta=|r-r_b|/\sqrt{E}$, we have

\[ (\na : \vu)^2= \frac{2}{E}\lac \lp\dzeta{u_\theta}\rp^2 +
\lp\dzeta{u_\varphi}\rp^2\rac_{r=r_b} \]

We now set $p=\cth -\omega$ and $q=\cth+\omega$. We thus get

\beqa (\na : \vu)^2= \frac{A^2 r_b^{2m}(\sth)^{2m-2}}{2E}\lac
(1+\cth)^2|p|e^{-\zeta\sqrt{2|p|}}\right. \\
+(1-\cth)^2|q|e^{-\zeta\sqrt{2|q|}}  \\
\left. -2\sin^2\theta\sqrt{pq}\;\Re e\lp
e^{2i\beta-\zeta(\sqrt{iq}+\sqrt{-ip})}\rp\rac \eeqa
Integrating over the $\varphi$-variable yields

\beqan \int_0^{2\pi} (\na : \vu)^2 d\varphi = \frac{\pi A^2
r_b^{2m}(\sth)^{2m-2}}{E}\hspace*{20mm}\nonumber \\
\times\lac
(1+\cth)^2|p|e^{-\zeta\sqrt{2|p|}}+(1-\cth)^2|q|e^{-\zeta\sqrt{2|q|}}\rac
\eeqan{eeqq}
We now integrate over the radial variable:

\beqa \int_\eta^1\int_0^{2\pi} (\na : \vu)^2 d\varphi\; r^2dr = \hspace*{20mm}\\
\frac{\pi A^2(\sth)^{2m-2}}{E} \int_\eta^1 r^{2m+2} f(\zeta) dr \eeqa
with
$f(\zeta)=(1+\cth)^2|p|e^{-\zeta\sqrt{2|p|}}+
(1-\cth)^2|q|e^{-\zeta\sqrt{2|q|}}$; since $r=\eta+\sqrt{E}\zeta$ or
$r=1-\sqrt{E} \zeta$ according to which side of the integral is chosen,
it turns out that

\beqa & &\hspace*{-15mm}\int_\eta^1\!\int_0^{2\pi} (\na : \vu)^2d\varphi\; r^2dr \\
&=& K\int_0^\infty \lc
(1+\cth)^2|p|e^{-\zeta\sqrt{2|p|}}\right. \\
&&+\left.(1-\cth)^2|q|e^{-\zeta\sqrt{2|q|}}\rc d\zeta \\
&=& \frac{K}{\sqrt{2}} \lc (1+\cth)^2\sqrt{|\cth-\omega|} \right. \\
&&+\left. (1-\cth)^2\sqrt{|\cth+\omega|} \rc
\eeqa
with $K=\pi A^2(\sth)^{2m-2}P(\eta)/\sqrt{E}$, where $P(\eta)$ is a
function depending on the boundary conditions (see table~\ref{bcp}). Finally
integrating over $\theta$, we find

\beq \int (\na : \vu)^2 dV = \frac{2\pi A^2P(\eta)}{\sqrt{2E}}
{\cal I}_m \eeqn{diss}
with

\beq  {\cal I}_m = \int_0^\pi
(1+\cth)^2\sqrt{|\cth-\omega|}\sin^{2m-1}\theta d\theta \eeq

Finally grouping \eq{ke} and \eq{diss}, we find the damping rate

\beq \gamma = -\frac{m(2m+3)!!}{2^{m+3/2}(m+1)!}Q(\eta)
{\cal I}_m\sqrt{E}
\eeqn{lf}
where $Q(\eta)=P(\eta)/\lp1-\eta^{2m+3}\rp$.

\begin{table}[t]
\centerline{
\begin{tabular}{ccc}
 Outer B.C. & no-slip & stress-free\\
 \hline \\
 Inner B.C. & & \\
no-slip & $1+\eta^{2m+2}$ & $\eta^{2m+2}$ \\
stress-free & 1 & No Ekman layer\\
\hline
\end{tabular}
}
\caption[]{Expression of the $P(\eta)$ function according to boundary
conditions.}
\label{bcp}
\end{table}

For the cases $m=1$ and $m=2$ we evaluated the expression of ${\cal
I}_m$, viz

\[ {\cal I}_1 = \frac{\sqrt{2}}{35}\lp 3^{5/2}+19\rp \]
\[{\cal I}_2 =
4\lp\frac{2}{3}\rp^{11/2}\frac{3401+2176\sqrt{2}}{5\times7\times9\times11}
\]
Other values are computed numerically and given in table~\ref{integ}.

The values given by \eq{lf} may be compared to other derivations, in
particular that of \cite{Green69} for $m=1$ who finds $\gamma/\sqrt{E} =
-2.62/\sqrt{2} = -1.8526$ ! For $m=2$, a direct numerical calculation,
similar to that of \cite{RV97}, gives $-2.482\sqrt{E}$ at $E=10^{-8}$
which is in good agreement with the analytical formula.

\begin{table}[t]
\centerline{
\begin{tabular}{ccccc} \hline
\\
 & $m=1$ & $m=2$ & $m=3$ & $m=4$ \\
\\
${\cal I}_m$ & 1.3976 & 0.80411 & 0.58075 & 0.46155 \\
${\gamma\over\sqrt{E}}$ & -1.8526 & -2.4876 & -3.0318 & -3.5339\\
\\
\hline
\end{tabular}
}
\caption[]{Values of the first integrals ${\cal I}_m$ and the
corresponding values of the scaled damping rates; this latter value
equals that of \cite{Green69} when multiplied by $\sqrt{2}$ because of
our choice of the time scale.}
\label{integ}
\end{table}

\section{Application to neutron stars and conclusions}

Let us now apply these results to the case of rapidly rotating neutron
stars. We take the viscosity from \cite{BU00}, $\nu=1.8\,f/T^2_8$ m$^2$/s
where $f$ is a dimensionless parameter taking into account the different
transport mechanisms in the fluid (superfluid phases for instance) and
$T_8$ is the temperature in 10$^8$K unit.
Using a radius of 12.53~km and an angular frequency of
$2\pi\times$1kHz, we find an Ekman number $\sim 10^{-12}$ which is indeed very
small and thus boundary layer theory applies.

\begin{figure}[t]
\centerline{\psfig{figure=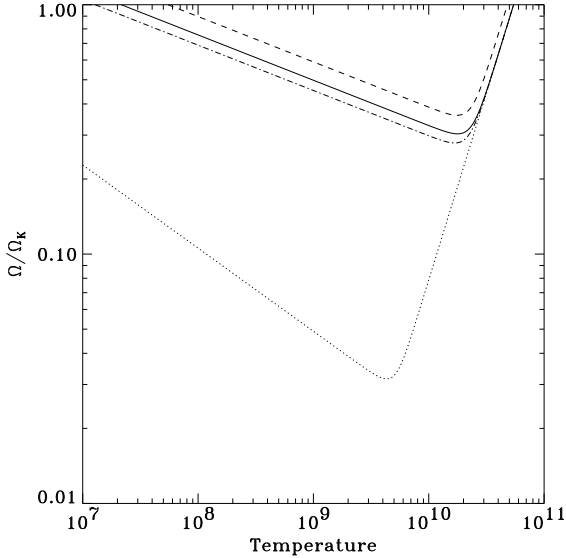,width=8cm}}
\caption[]{Curves of critical angular velocity, normalized by
$\Omega_K=\frac{2}{3}\sqrt{\pi G\overline{\rho}}$, for different models.
The solid line shows the result of the present work, the dashed-dotted
one is that of \cite{AJKS00} and the dashed one is for \cite{BU00}. The
dotted line is the critical curve for a non-crusted star. No core has been
included ($\eta=0$).}
\label{crit} \end{figure}

We may now estimate the charateristic time scale for the damping of the
$m=2$-mode. We find

\beq T_{d} = 26.7 {\rm s}\; \frac{T_8}{\sqrt{f}}\lp\frac{R}{10\;\rm
km}\rp \lp\frac{1\;\rm kHz}{\nu_s}\rp^{1/2} \eeq
which is a somewhat smaller value than the previous estimate of
\cite{BU00} and \cite{AJKS00} who find a characteristic time of 100~s
and 200~s respectively. Our disagreement with these authors comes
from their approximate evaluation of the boundary layer dissipation and
from the resulting functional dependence with respect to mass and
density. Let us first evaluate the damping rate according to
\cite{LL7189}; it turns out that

\beq 2\gamma = -\lp\frac{\omega E}{2}\rp^{1/2}\frac{\lp \int_{4\pi}
\vu^2\sth d\theta d\varphi\rp(r=1)}{\intvol \vu^2 dV} \eeq
where we used our non-dimensional units. Since the radial dependence of
the modes is in $r^m$ and $\omega=1/(m+1)$, we easily find that

\[ \gamma = -\frac{2m+3}{2\sqrt{2m+2}} \sqrt{E} \]
When this expression is applied to the $m=2$-mode, we find that
$\gamma=-1.429\sqrt{E}$ which is a factor $1.74$ weaker than the correct
result. 

If we use, as previous authors, a step function for describing the
density difference between that of the crust and the mean density, we
find that the damping rate reads

\beq \gamma_{Ek} = -2.4876\sqrt{E}\frac{\rho_b}{\overline{\rho}}\,
2\Omega = -0.0346 \frac{\rho_b}{\overline{\rho}}
\frac{\sqrt{f\Omega_\star}}{T_8} {\rm s}^{-1}\eeqn{damp_rate}
where $\rho_b$ is the density of the fluid just below the crust and
$\Omega_\star=\Omega/\sqrt{\pi G\overline{\rho}}$.

Our calculation therefore shows that the window
of instability in the $\Omega, T$ plane is smaller than previously
estimated for crusted neutron stars.

Considering a 1.4~M${_\odot}$ neutron star with a radius of 12.53~km as a
test case, the growth rate of the mode due to gravitational radiation is
$\gamma_{gw} = 0.658 {\rm s}^{-1} \Omega_\star^6$
\cite[we use the expression given in][]{LOM98}; although, it is not
relevant for an incompressible fluid, we take into account the damping
rate due to bulk viscosity in order to ease comparison with previous
work; from \cite{LMO99}, we find $\gamma_{bulk} = -2.2\, 10^{-12}\;{\rm
s}^{-1}\; T_9^6\Omega_\star^2$. From \eq{damp_rate}, we have
$\gamma_{Ek}= -1.53\, 10^{-3}\;{\rm s}^{-1}\; \Omega_\star^{1/2}/T_9$ where
we took $\rho_b=1.5\,10^{17}$kg/m$^3$; solving the equation

\[ \gamma_{gw}+\gamma_{Ek}+\gamma_{Bulk} = 0 \]
for different values of the temperature yields the curves displayed in
figure~\ref{crit}.

As expected, we see that the window of instability narrows compared to
\cite{AJKS00}: for a given temperature, the critical angular velocity
raises by $\sim$10\% typically.

Another interesting conclusion of this work is that the presence of a
solid inner core does not change the damping rates very much unless its
radius is close to unity. The reason for that is to be found in the
shape of the inertial r-modes whose amplitudes are concentrated near the
outer boundary. Therefore, the rotating instability of rapidly rotating
stars is quite insensitive to the presence of a solid core and more
generally to any phase transition which does not occur close to the
surface.

\acknowledgments

I am very grateful to S.~Bonazzola and E.~Gourgoulhon for drawing my
attention on these questions and for helpful discussions. I am also
very grateful to Ian Jones for his note about the density profile used
in models of neutron stars in previous work.  Part of the calculations
have been carried out on the Nec SX5 of IDRIS at Orsay and on the
CalMip machine of CICT in Toulouse which are gratefully acknowledged.

\bibliography{../../biblio/bibnew}

\end{document}